\documentclass[aps,prd,preprintnumbers,twocolumn,amsmath,amssymb,showpacs,floatfix]{revtex4}
\usepackage{epsfig}
\usepackage{dcolumn}
\usepackage{bm}

\begin{document}

\title{Photon Radiation and Dilepton Production\\
        Induced by Rescattering in Strong Interacting Medium}

\author{Hanzhong Zhang, Zhongbo Kang, Ben-Wei Zhang and Enke Wang}

\affiliation{Institute of Particle Physics, Huazhong Normal
University, Wuhan 430079, China; Key Laboratory of Quark $\&$ Lepton
Physics (HZNU), MOE, China}


\begin{abstract}
Using the opacity expansion technique we investigate the photon
radiation and dilepton production induced by multiple rescattering
as energetic parton jet passing through the strong interacting
medium. The real photon radiation and dilepton invariant-mass
spectrums of the bremsstrahlung contribution from an energetic
quark jet are presented. The leading contribution of total energy loss
by photon emmision in medium of a higher energy quark jet
is found to be proportional to
the jet energy and has a linear dependence on the thickness of the
nuclear target. The rescattering contribution for the dilepton is
important only for small value of the invariant-mass and for the
not so fast jet. The contribution fraction of the dilepton induced by
rescattering in medium is found to be nearly a constant when the ratio
of jet energy to Debye screening mass $E/\mu$ is large.
\end{abstract}


\maketitle


\section{Introduction}

Hard processes are considered as a good tool to study the properties
of the quark matter produced in ultra-relativistic heavy-ion
collision because it can probe the early stage of the evolution of
dense system, during which a quark-gluon plasma (QGP) could exist
for a short period of time. One important aspect of hard processes
is jet's energy loss or jet quenching
\cite{Wang94,Baier95,Zakharov96,Gyulassy00,Wiedemann01,Wang01,Wang02,xnw04,zoww07}
due to gluon radiation induced by multiple rescattering as energetic
parton jet going through the strong interacting medium. Jet
quenching in high energy nucleus-nucleus collisions is predicted to
lead to strong suppression of both single- and correlated away-side
dihadron spectra at large transverse momentum as compared to $p+p$
collisions at the same energy~\cite{Wang94,xnw04,zoww07}. These
phenomenon have been observed in central nucleus-nucleus collisions
in recent RHIC experiments
\cite{phenix02,phenix03,star02,star03a,star03b,phenix-star-single,star03-06}.
A simultaneous $\chi^2$-fit to both single and dihadron spectra can
be achieved within their minima in the same narrow range of energy
loss parameter for two different measurements at RHIC~\cite{zoww07}.
This fact provides convincing evidence for jet quenching
description.

In addition to the gluon radiation, an energetic quark jet suffering
multiple rescattering in medium should also induce real and virtual
photon bremsstrahlung, where the emitted virtual photon may further
decay into a dilepton. Since the (real or virtual) photon interacts
with the particles in the collision region only through the
electromagnetic coupling, its mean-free path is expected to be quite
large\cite{Wong-book}. One expects the photons can pass through the
collision region without rescattering and then carry the information
of medium at the time they have been produced, which may help us
analyze the properties of the hot nuclear matter created in
relativistic nucleus-nucleus collisions and provide complementary
test of jet quenching mechanism due to gluon emission off their same
parent jet\cite{JOS,JJS,Zakharov04}.

The bremsstrahlung photon production in nucleus-nucleus collisions
has been previously discussed by many researchers. B. G. Zakharov
studied\cite{Zakharov04} the induced photon bremsstrahlung from a
fast quark produced in $A+A$ collisions due to multiple scattering
in QGP and predicted the medium- induced photon emission may
enhanced the photon production at high $p_T$ in $A+A$ collisions by
$30\%$ as compared to $p+p$ collisions, which is not consistent with
the recent PHENIX measurements \cite{Isobe:2007ku}. By using
diagrammatic method in thermal field theory, Arnold, Moore and Yaffe
(AMY)\cite{AMY} calculated the photon and gluon emission rates of an
equilibrated, hot QCD plasma to the leading order in both $\alpha_e$
and the QCD coupling $g_s(T)$, where the energy loss of an
asymptotic jet in QGP is investigated by assuming a very high
temperature to guarantee $g_s(T)\ll1$. Later S. Turbide, C. Gale, S,
Jeon and G. Moore (TGJM)\cite{TGJM} utilize AMY formalism to study
high $p_T$ pion and photon production including bremsstrahlung
process in heavy-ion collisions at RHIC and LHC. However, their
results overestimate the photon production at $Au+Au$ with
$\sqrt{s_{NN}}=200$~GeV measured by PHENIX experiments
\cite{Isobe:2007ku}.

Therefore, it is interesting and beneficial to make an complementary
calculation of photon production in $A+A$ collision with different
approaches. In this paper we follow the framework of opacity
expansion developed by Gyulass, Levai and Vitev (GLV) in
Ref.\cite{Gyulassy00} to study the bremsstrahlung photon and
dilepton production induced by multiple rescattering as the parton
jet propagating in the strong interacting medium. The GLV formalism
has been extensively applied to study the high $p_T$ phenomenon at
RHIC and the phenomenological studies based on GLV formalism have
described different measurements at RHIC very well
\cite{Gyulassy:2000gk,Vitev:2002pf,Vitev:2005he}. In GLV formalism
the contributions of radiative energy loss in expanded in powers of
the opacity, which is defined as the averaged scattering number,
$L/\lambda$ with $L$ representing the thickness of the target and
$\lambda$ the gluon mean-free path. And it has been shown that the
opacity expansion series converge very rapidly and the first order
contribution is dominant. For the comparison of several jet
quenching approaches including AMY and GLV formalism, please see the
review in Ref.\cite{Majumder:2007iu}.

As a first step we concentrate on the production of photon and
dilepton emitted off a fast quark jet at the first order opacity. In
our calculation the QCD coupling $g_s$ from the scattering point
where a quark jet interacts with a parton in a QGP system, is put in
a potential function and absorbed into opacity. Therefore, compared
with AMY's study\cite{AMY}, our result of photon emission rate at
first order opacity is also to the leading order in both $\alpha_e$
and $\alpha_s$. Our conclusions about bremsstrahlung photon in
strong interacting medium are qualitatively consistent with AMY's
studies. We confirm that the leading contribution of energy loss by
photon emission has a linear dependence on the thickness of the
nuclear target $L$\cite{bwz-w}, which is different from the energy
loss by gluon radiation where $\Delta E$ quadratically depend on the
thickness $L$. We observe that the rate of photon radiation as a
function of energy fraction $x$ carried by radiated photon has a
peak at small $x \sim 0.3$, whereas in Zakharov's study the peak of
the rate by induced photon correction is at $x \sim 0.9$. This may
explain why a larger enhancement factor of photon production at high
$p_T$ in $A+A$ is found in Zakharov's study. For a jet created in
$A+A$ collisions, multiple scattering induces photon radiation as
well as gluon radiation (jet quenching). Compared with the case of
$p+p$ collisions, induced photon radiation enhances the photon yield
while jet quenching decreases the photon yield from jet
fragmentation in $A+A$ collisions\cite{JOS,JJS,Zakharov04}. If the
former effect is overestimated larger than the latter effect, the
nuclear modification factor for medium and large $p_T$ photon
production $R_{AA}>1$ will appear in $A+A$ collisions, not
consistent with the recent PHENIX measurements \cite{Isobe:2007ku}.

For dilepton production it is shown the rescattering contribution is
important only for small value of the invariant-mass and also for
the not so fast jet, and the contribution fraction of dilepton
induced by multiple scattering to approach a constant when the quark
jet energy $E/\mu
>30$, where $E$ denotes the energy of the jet, and $\mu$ the debye
screening mass.

This paper is organized as follows. In Sec. II we present the
calculation of the jet energy loss by photon emission in hot medium
and the resulting photon spectrum due to real photon radiation. In
Sec. III we investigate the invariant-mass spectrum in the dilepton
production due to virtual photon radiation. Corresponding numerical
analysis are given in these two sections, respectively. In Sec. IV
we make a brief conclusion.

\section{Induced real photon radiation}

Consider a source localized at $y_0=(t_0, {\bf y}_{0})$ that
produces a parton jet described by a wave packet $J(p)$. The
parton jet undergoes multiple rescattering by exchanging a gluon
with a target parton and emits a photon with light-cone 4-momentum
and polarization,
\begin{eqnarray}
   k&=&\left[xp^+, \frac{{\bf k}^2_{\perp}}{xp^+}, {\bf k}_{\perp}\right]\, ,
\label{k}
\\
  \epsilon&=&\left[0, \frac{{2\bf \epsilon}_{\perp}\cdot{\bf k}_{\perp}}{xp^+},
            {\bf\epsilon}_{\perp}\right]\, ,
\label{eps}
\end{eqnarray}
where $x$ is the energy fraction carried by photon from the jet
parton. The initial and final four momentum of the jet is assumed
as
\begin{eqnarray}
   p_i&=&[p^+,p^-,0_{\perp}],\\
   p_f&=&\left[(1-x)p^+, \frac{({\bf q}_{\perp}-{\bf k}_{\perp})^2+m^2}
   {(1-x)p^+}, {\bf q}_{\perp}-{\bf k}_{\perp}\right]\, ,
\label{pipf}
\end{eqnarray}
where ${\bf q}_{\perp}$ and $m$ denote the transverse momentum
transfer and thermal mass of the jet parton in the hot medium,
respectively.

As proposed by Gyulassy and Wang\cite{Wang94}, the interaction
between the jet and the target parton can be modeled by the static
color-screened Yukawa potential,
\begin{eqnarray}
  V_n&=&2\pi\delta(q^0)v({\bf q}_n)e^{-i{\bf q}_n\cdot{\bf y}_n}T_{a_n}(j)
  \otimes T_{a_n}(n),\\
  v({\bf q})&=&\frac{4\pi\alpha_s}{{\bf q}_n^2+\mu^2}\, ,
\label{vn}
\end{eqnarray}
where $\alpha_s$ is the strong coupling constant, $\mu$ the Debye
thermal mass of gluon in the hot medium, ${\bf q}_n$ the momentum
transfer from a target parton $n$ at the coordinates ${\bf
y}_n=(z_n, {\bf y}_{\perp n})$. $T_{a_n}(j)$ and $T_{a_n}(n)$ are
the color matrices for the jet and target parton, respectively. As
assumed in Ref.\cite{Gyulassy00}, all target partons are in the same
$d_T$ dimensional representation with Casimir $C_2(T)$ ($Tr
T_a(n)=0$ and
$Tr(T_a(i)T_b(j))=\delta_{ij}\delta_{ab}C_2(i)d_T/d_A$). For color
matrices generators of the jet in the $d_R$ dimensional
representation,  $Tr(T_a(j)T_a(j)) = C_R d_R$.

 \begin{figure}
 \vskip -.5cm
 \epsfxsize 120mm \epsfbox{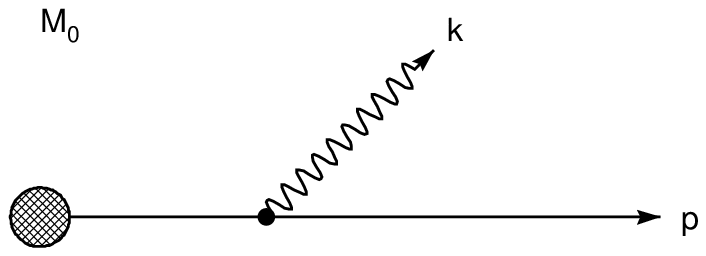}
 \vskip -1.5cm
 \epsfxsize 90mm \epsfbox{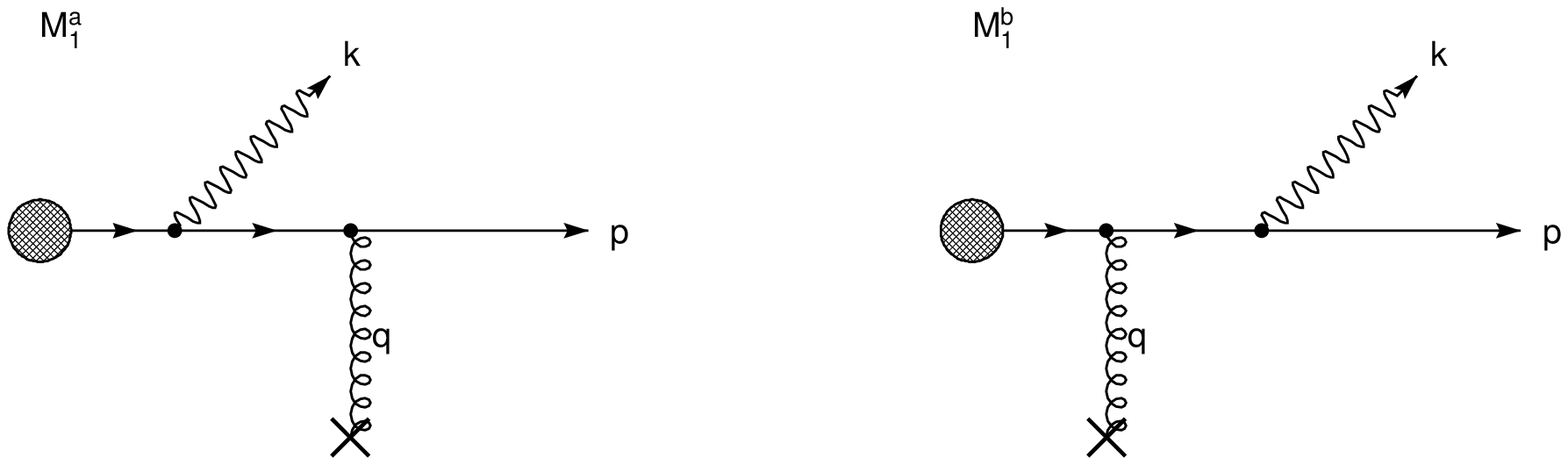}
 \vskip 0.1cm
 \epsfxsize 90mm \epsfbox{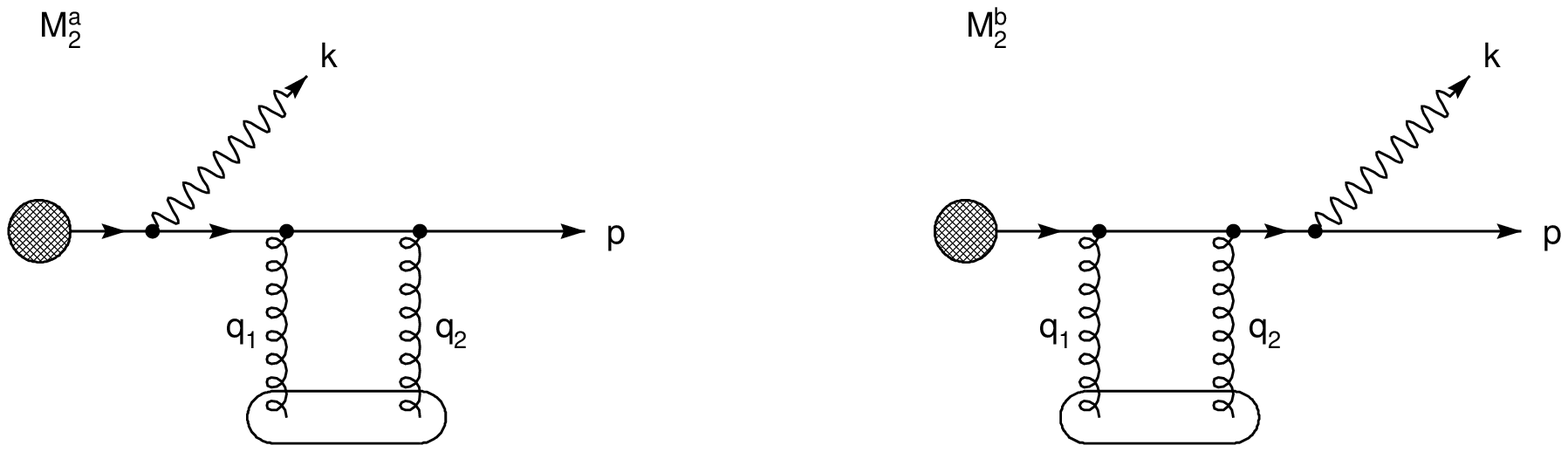}
 \vskip 0.2cm
 \caption{\label{fig1}
 \small Feynman diagram for photon radiation induced by
   self-quenching, single and double
   Born rescattering in the medium.}
 \end{figure}

As shown in Fig.1, we denote ${\cal M}_0$, ${\cal M}_1$ and ${\cal
M}_2$ as the scattering amplitudes for self-quenching, single
rescattering and the double Born rescattering, respectively. The
squared total amplitude is given by
\begin{eqnarray}
  |{\cal M}_s|^2&=&|{\cal M}_0+{\cal M}_1+{\cal M}_2+\cdots|^2
  \nonumber\\
  &=&|{\cal M}_0|^2+|{\cal M}_1|^2
  +2Re({\cal M}_1{\cal M}_0^*)
  \nonumber\\
  &&+2Re({\cal M}_2{\cal M}_0^*)+\cdots\, .
\label{Ms}
\end{eqnarray}

For the self-quenching shown in Fig.1 we get
\begin{eqnarray}
  {\cal M}_0=iJ(p+k)e^{i(p+k)\cdot y_0}{\cal R}_0\, ,
\label{M0}
\end{eqnarray}
where the radiation amplitude ${\cal R}_0$ can be expressed as
\begin{eqnarray}
  {\cal R}_0=g\frac{\epsilon\cdot k}{p\cdot k}={2g(1-x)}
   \frac{{\bf\epsilon}_{\perp}\cdot{\bf k}_{\perp}}
   {{\bf k}_{\perp}^2+x^2m^2}\, ,
\label{R0}
\end{eqnarray}
$g$ is the coupling constant between quark and photon.

For multiple rescattering we assume the packet $J(p)$ vary slowly
over the range of momentum transfers supplied by the potential,
the rescattering centers are well separated and the distance
between two successive scattering is large compared to the
interaction range,
\begin{equation}
   z_i-z_{i-1}\gg 1/\mu\, .
\label{sep}
\end{equation}

For single scattering illustrated in Fig.1 the rescattering
amplitude of the first Feynman diagram can be expressed as
\begin{eqnarray}
  {\cal M}_1^a&=&iJ(p+k)e^{i(p+k)\cdot y_0}
    \sum_{j=1}^N\int\frac{d^2{\bf q}_{\perp}}{(2\pi)^2}
    e^{-i{\bf q}_{\perp}\cdot({\bf y}_{\perp j}-{\bf y}_{0\perp})}
 \nonumber\\
    &&\times\int\frac{dq_z}{2\pi}
     \frac{2g(1-x)p^+\epsilon\cdot p_i}
     {[(p-q)^2+i\epsilon][(p+k-q)^2+i\epsilon]}
 \nonumber\\
     &&\times v(q_z , {\bf q}_{\perp})e^{-iq_z(z_j-z_0)}
     T_a T_a(j)\, ,
\label{M1a}
\end{eqnarray}
where $N$ is the number of the targets in the medium. The
integration over $q_z$ is performed by closing the contour below
the real axis in the complex $q_z$ plane. Two poles
${q_z}'=-i\epsilon$ and ${q_z}''=-\omega_0-i\epsilon$ contribute
to the scattering amplitude. Here $\omega_0$ can be written as
\begin{eqnarray}
   \omega_0=\frac{{\bf k}_{\perp}^2+x^2m^2}{xp^+}\, .
\label{om0}
\end{eqnarray}
The contribution from the pole corresponding to the potential
singularities can be neglected because of the suppression factor
$\exp[-\mu(z-z_0)]$. From Eq.(\ref{M1a}) we deduce the radiation
amplitude for first Feynman diagram of the single rescattering in
Fig.1 as
\begin{equation}
   {\cal R}_1^a=-i2g(1-x)\frac{\epsilon_{\perp}\cdot {\bf k}_{\perp}}{{\bf k}^2_{\perp}+x^2m^2}
   \sum_{j=1}^N\left[1
   -e^{i\omega_0(z_j-z_0)}\right]\, .
\label{R1a}
\end{equation}
Similarly we get the radiation amplitude for second Feynman
diagram of the single rescattering in Fig.1 as
\begin{eqnarray}
   {\cal R}_1^b&=&-i2g(1-x)\frac{\epsilon_{\perp}\cdot ({\bf k}_{\perp}-x{\bf q}_{\perp})}
   {({\bf k}_{\perp}-x{\bf q}_{\perp})^2+x^2m^2}
 \nonumber\\
   &&\times\sum_{j=1}^N e^{i\omega_0(z_j-z_0)}\, .
\label{R1b}
\end{eqnarray}
Then the total radiation amplitude for single scattering is given
by
\begin{eqnarray}
   {\cal R}_1&=&{\cal R}_1^a+{\cal R}_1^b=-i2g(1-x)
 \nonumber\\
   &\times&{\bf\epsilon}_{\perp}\cdot \sum_{j=1}^N\left[{\bf B}
   -\left({\bf B}-{\bf C}\right)
  e^{i\omega_0(z_j-z_0)}\right]\, ,
\label{R1}
\end{eqnarray}
where
\begin{eqnarray}
   {\bf B}=\frac{{\bf k}_{\perp}}{{\bf k}_{\perp}^2+x^2m^2},
   \quad{\bf C}=\frac{{\bf k}_{\perp}-x{\bf q}_{\perp}}
   {({\bf k}_{\perp}-x{\bf q}_{\perp})^2+x^2m^2}\, .
\label{B}
\end{eqnarray}

For the double Born virtual interaction shown in Fig.1,
$y_1=y_2=y_j$, the rescattering amplitude ${\cal M}_2^a$ reads
\begin{eqnarray}
    {\cal M}_2^a&=&iJ(p+k)e^{i(p+k)\cdot y_0}\sum_{j=1}^N
    \int\frac{d^2{\bf q}_{1\perp}}{(2\pi)^2} \int\frac{d^2{\bf q}_{2\perp}}{(2\pi)^2}
  \nonumber\\
    &\times& e^{-i({\bf q}_{1\perp}+{\bf q}_{2\perp})\cdot
    ({\bf y}_{\perp j}-{\bf y}_{\perp 0})}(1-x)^2{p^+}^2\int\frac{dq_{2z}}{2\pi}
  \nonumber\\
    &\times& v({\bf q}_2)T_{a_2}T_{a_2}(j)\Delta(p-q_2)\int\frac{dq_{1z}}{2\pi}v({\bf q}_1)T_{a_1}T_{a_1}(j)
  \nonumber\\
    &\times& e^{-i(q_{1z}+q_{2z})(z_j-z_0)}2g[\epsilon\cdot(p-q_1-q_2)]
  \nonumber\\
    &\times& \Delta(p-q_1-q_2)\Delta(p+k-q_1-q_2)\, .
\label{R2a}
\end{eqnarray}
Integrate over $q_{1z}$ by closing the contour below the real axis
in the complex $q_{1z}$ plane, the residue of the poles
$q_{1z}^{'}=-q_{2z}-i\epsilon$ and
$q_{1z}^{''}=-q_{2z}-\omega_0-i\epsilon$ contribute to the
integral. The contribution from the potential singularities can be
ignored because $z_j-z_0\gg 1/\mu$. so ${\cal M}_2^a$ can be
rewritten as
\begin{eqnarray}
    {\cal M}_2^a&=&iJ(p+k)e^{i(p+k)\cdot y_0} \sum_{j=1}^N
    \int\frac{d^2{\bf q}_{1\perp}}{(2\pi)^2} \int\frac{d^2{\bf q}_{2\perp}}{(2\pi)^2}
  \nonumber\\
    &\times& e^{-i({\bf q}_{1\perp}+{\bf q}_{2\perp}) \cdot ({\bf y}_{\perp j}-{\bf y}_{\perp 0})}
    (1-x)^2{p^+}^2\int\frac{dq_{2z}}{2\pi}
  \nonumber\\
    &\times& \frac{-2ig}{\omega_0(1-x){p^+}^2}v(q_{2z}, {\bf q}_{2\perp})
    v(-q_{2z}, {\bf q}_{1\perp})\Delta(p-q_2)
  \nonumber\\
    &\times& \epsilon\cdot p_i
    \left[1-e^{i\omega_0(z_j-z_0)}\right]T_{a_1}T_{a_1}(j)T_{a_2}T_{a_2}(j)\, .
\label{R2a'}
\end{eqnarray}
For $q_{2z}$ integration, there is no exponentially suppressed
factor, so the potential singularities $q_{2z}=-i\mu_1$ and
$-i\mu_2$ contribute to the integral. By closing the integration
contour below the real axis in the complex $q_{2z}$ plane and
taking the residue in the poles $q_{2z}^{'}=-i\epsilon$ and
$q_{2z}^{''}=-i\sqrt{{\bf q}_{2\perp}^2+\mu^2}$, we get the
radiation amplitude
\begin{equation}
  {\cal R}_2^a(Born)=-g(1-x){\bf\epsilon}_{\perp}\cdot{\bf B}
  \sum_{j=1}^N\left[1-e^{i\omega_0(z_j-z_0)}\right].
\label{R2a''}
\end{equation}

In a similar way, from the last Feynman diagram in Fig. 1 we
obtain the rescattering amplitude ${\cal M}_2^b$. The
corresponding radiation is given by
\begin{equation}
   {\cal R}_2^b(Born)=-g(1-x){\bf\epsilon}_{\perp}\cdot{\bf B}
   \sum_{j=1}^N e^{i\omega_0(z_j-z_0)}\, .
\label{R2b}
\end{equation}

The total radiation amplitude from the double Born virtual
interaction can be expressed as
\begin{eqnarray}
     {\cal R}_2^B&=&{\cal R}_2^a(Born)+{\cal R}_2^b(Born)
  \nonumber\\
     &=&-g(1-x)N{\bf\epsilon}_{\perp}\cdot{\bf B}\, .
\label{R2}
\end{eqnarray}

As shown in Ref.\cite{Gyulassy00}, in opacity expansion technique
we make following simplification: define the relative transverse
coordinate ${\bf b}_{\perp j}={\bf y}_{\perp j}-{\bf y}_{\perp
0}$, it varies over a large transverse area $A_{\perp}$. Then the
ensemble average over the rescattering center location reduces to
an impact parameter average,
\begin{equation}
  \langle\dots\rangle=\frac{1}{A_{\perp}}\int d^2{\bf b}_{\perp} dz_j dz_0\rho(z_j,z_0)\dots
\label{average}
\end{equation}

Along the longitudinal direction, the distribution density for the
locations of the rescattering center is defined by
\begin{eqnarray}
     \rho(z_0,z)&=&\left[\frac{\theta(L-z)}{L/2}Exp\left(-\frac{L-z}{L/2}
     \right)\right]
  \nonumber\\
     &\times&\left[\frac{\theta(z-z_0)}{L/2}Exp\left(-\frac{z-z_0}{L/2}\right)\right]\, .
\label{rho}
\end{eqnarray}

Taking the initial color average and the ensemble average we
obtain
\begin{eqnarray}
    \langle|{\cal M}_1|^2\rangle&=&\int dz_0 dz\rho(z_0,z)
    \frac{N\sigma_{el}}{A_{\perp}}
  \nonumber\\
    &\times&\int d^2{\bf q}_{\perp}\frac 1{\sigma_{el}}
     \frac{C_RC_2(T)}{d_A}\frac{|v(0,{\bf q}_{\perp})|^2}{(2\pi)^2}|{\cal R}_1|^2
\label{M1^2}
\end{eqnarray}

The elastic cross section with the small transverse momentum
transfer between the jet and target partons
is\cite{Wang94,Gyulassy00}
\begin{equation}
   \frac{d\sigma_{el}}{d^2{\bf q}_{\perp}}
   =\frac{C_RC_2(T)}{d_A}\frac{|v(0,{\bf q}_{\perp})|^2}{(2\pi)^2}.
\label{sigma_el}
\end{equation}
One can defining $|\bar v(0,{\bf q}_{\perp})|^2$ as the normalized
distribution of momentum transfers from the scattering
centers\cite{Gyulassy00},
\begin{equation}
   \frac1{\sigma_{el}}\frac{d\sigma_{el}}{d^2{\bf q}_{\perp}}
   \equiv|\bar v(0,{\bf q}_{\perp})|^2
    =\frac1\pi\frac{\mu^2_{eff}}{({\bf q}_{\perp}^2+\mu^2)^2}\, ,
\label{sigma_el'}
\end{equation}
where
\begin{eqnarray}
     \frac1{\mu_{eff}^2}=\frac1{\mu^2}-\frac1{\mu^2+
     {\bf q}_{\perp}^2}\, ,
\label{mu_eff}
\end{eqnarray}
to insure $\int^{{\bf q}_{\perp max}}d^2{\bf q}_{\perp}|\bar
v(0,{\bf q}_{\perp})|^2=1$. In numerical estimation, we take that
${\bf q}^2_{\perp max}=\infty$, $\mu_{eff}\approx\mu$.

Using the normalized distribution $|\bar v(0,{\bf q}_{\perp})|^2$,
we can rewrite Eq.(\ref{M1^2}) as
\begin{eqnarray}
    \langle|{\cal M}_1|^2\rangle&=&\frac{L}{\lambda}\int d^2{\bf q}_{\perp}|
    {\bar v}(0,{\bf q}_{\perp})|^2
  \nonumber\\
    &\times&\int dz_0 dz\rho(z_0,z)|{\cal R}_1|^2\, ,
\label{M1^2'}
\end{eqnarray}
where
\begin{equation}
   \frac{L}{\lambda}=\frac{N\sigma_{el}}{A_{\perp}}
   \equiv{\bar n}\, .
\label{opacity}
\end{equation}
${\bar n}$ represents the mean number of the rescattering which is
defined as the so-called opacity\cite{Gyulassy00}.

The first interference term $Re({\cal M}_1{\cal M}_0^*)$ in
Eq.(\ref{Ms}) doesn't contribute to photon radiation probability
because of $Tr(T_a(j))=0$. The contribution from the double Born
virtual interaction is given by
\begin{eqnarray}
    \langle2Re({\cal M}_2^B{\cal M}_0^{*})\rangle&=&\frac{L}{\lambda}\int d^2{\bf q}_{\perp}
    |\bar v(0,{\bf q}_{\perp})|^2
  \nonumber\\
    &\times& \int dz_0 dz\rho(z_0,z)2Re({\cal R}_2^B{\cal
    R}_0^{*})\, .
\label{M2M0}
\end{eqnarray}

Denote $N$ as the number of the radiated photon, we have
\begin{equation}
    dN=\langle|{\cal M}|^2\rangle \frac{d^3{\bf k}}{(2\pi)^3 2\omega}\, .
\label{dN}
\end{equation}
The transverse momentum spectrum of the radiated photon can be
expressed as
\begin{equation}
    \frac{dN}{d|{\bf k}_{\perp}|^2dx} =\frac{1}{4(2\pi)^2}\frac{1}{x}
    \langle|{\cal M}|^2\rangle\, .
\label{dNdk}
\end{equation}

To the first order of opacity expansion, the square of the
rescattering amplitude is given by
\begin{eqnarray}
    \langle|{\cal M}|^2\rangle&=&\langle|{\cal M}_0|^2+|{\cal M}_1|^2
    + 2Re({\cal M}_2^B{\cal M}_0^{*})\rangle
  \nonumber\\
    &=&|{\cal R}_0|^2+\frac{L}{\lambda} \int\frac{\mu^2d^2{\bf q}_{\perp}}
    {\pi({\bf q}^2_{\perp}+\mu^2)^2}\int dz_0 dz
  \nonumber\\
    &\times&\rho(z_0,z)[|{\cal R}_1|^2+ 2Re({\cal R}_2^B{\cal R}_0^{*})],
\label{M^2}
\end{eqnarray}
where
\begin{eqnarray}
    |{\cal R}_1|^2&+&2Re({\cal R}_2^B{\cal R}_0^{*})
    =4g^2(1-x)^2[({\bf B}-{\bf C})^2
  \nonumber\\
    &-&2\cos(\omega_0(z-z_0)) ({\bf B}^2-{\bf B}\cdot{\bf C})]\, .
\label{R}
\end{eqnarray}
The term with $\cos(\omega_0(z-z_0))$ reflects the destructive
interference arising from the Abelian LPM effect\cite{LPM}. The
integration of this term over target distribution gives
\begin{eqnarray}
    I(\omega_0,L)&\equiv &
    \int dz_0dz\rho(z_0,z)\cos(\omega_0(z-z_0))
  \nonumber\\
    &=&\frac{4}{4+\omega_0^2L^2}\, .
\label{lpm}
\end{eqnarray}

Denote $dN_{0+1}=dN_0+dN_1$, $N_0$ is the number of radiated
photon at zero-order opacity expansion corresponding to
self-quenching term $|{\cal M}_0|^2$, $N_1$ the number of radiated
photon at first-order opacity expansion corresponding to the sum
of single rescattering term $|{\cal M}_1|^2$ and the interference
term between self-quenching and double born virtual interaction
$2Re({\cal M}_2^B{\cal M}_0^*)$. Defining
\begin{equation}
   u=|{\bf q}_{\perp}|^2/\mu^2\, ,
   \quad y=|{\bf k}_{\perp}|^2/\mu^2\, ,
   \quad w=m^2/\mu^2\, ,
\label{uyw}
\end{equation}
we obtain
\begin{equation}
   \frac{dN_{0}}{dxdy}= \frac{4\alpha_e}{9\pi}\frac{(1-x)^2}{x}
   \frac{y}{(y+x^2w)^2}\, ,
\label{dN0/dxdy}
\end{equation}

\begin{equation}
   \frac{dN_{1}}{dxdy}=
   \frac{4\alpha_e}{9\pi}\frac{(1-x)^2}{x}\frac{L}{\lambda}\int
   du\frac1{(1+u)^2}f(x,u,y)\, ,
\label{dN1/dxdy}
\end{equation}
where function $f(x,u,y)$ is given by
\begin{eqnarray}
    f(x,u,y)=\frac{(y+x^2u)(y+x^2w+x^2u)-4x^2uy}
    {\sqrt{((y+x^2u+x^2w)^2-4x^2uy)^3}}
  \nonumber\\
    \quad-\frac{y}{(y+x^2w)^2}+
    \left(1-I(\omega_0,L)\right)\frac{1}{y+x^2w}
  \nonumber\\
    \quad\times\left( \frac{y-x^2w}{y+x^2w}
    -\frac{y-x^2w-x^2u}{\sqrt{(y+x^2u+x^2w)^2-4x^2uy}}
    \right )\, .
\label{f(xuy)}
\end{eqnarray}
In getting above expressions the integral of the angle between
${\bf k}_{\perp}$ and ${\bf q}_{\perp}$ has been integrated out
and we have chosen the coupling constant $g=2e/3$ for $u$ light
quark jet.

If setting the mass of quark to vanish, from Eq.(\ref{uyw}) we see
$w=0$, then function $f(x,u,y)$ in Eq.(\ref{f(xuy)}) reduce to
\begin{eqnarray}
    f(x,u,y)&=&\frac1{|y-x^2u|}-\frac{1}{y}
  \nonumber\\
    &+&\left(1-I(\omega_0,L)\right)\frac{1}{y}
    \left(1-\frac{y-x^2u}{|y-x^2u|}\right)\,.
\label{f'(xuy)}
\end{eqnarray}
This function have two collinear divergences. One is from the
initial-state bremsstrahlung as $y\sim 0$ (or ${\bf k}_{\perp}\sim
0$) along the initial jet direction; the another one is from the
final-state bremsstrahlung along the final jet direction as $|{\bf
k}_{\perp}|\sim x|{\bf q}_{\perp}|$ (or emission angle $\theta_e$
is near to the scattering angle $\theta_s$), here $\theta_e$ and
$\theta_s$ approximate
\begin{equation}
  \theta_e\sim \frac{|{\bf k}_{\perp}|}{k^+}\, ,
  \qquad \theta_s\sim \frac{|{\bf q}_{\perp}|}{p^+}\, .
\label{theta_sa}
\end{equation}
Above two collinear divergences is caused by setting quark jet's
mass to be zero. As $\theta_e > \theta_s$, the third term in
Eq.(\ref{f'(xuy)}) vanish, so there is no interference. In purely
QED case, as $\theta_e < \theta_s$ the radiative energy loss of the
charged particle has been studied in Ref.\cite{Baier96} in which the
charged particle undergoes multiple rescattering by exchanging
photon. Here we investigate the transverse momentum spectrum of
radiated photon and the corresponding energy loss of energetic quark
jet which undergoes multiple rescattering by exchanging gluon.

 \begin{figure}
 \epsfxsize 85mm \epsfbox{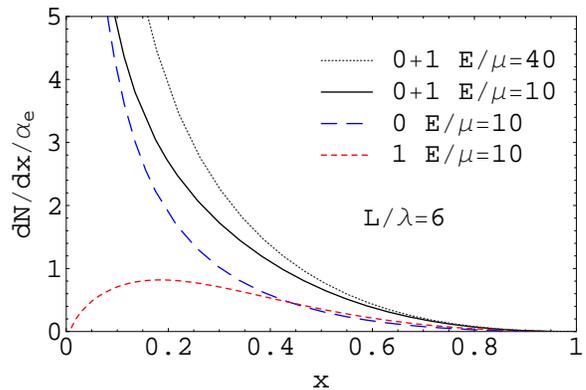}
 \vskip 0.2cm
 \caption{\label{fig2}
 \small (Color online) The production rates of the photon emitted off
 two quark jets with energy $E/\mu=10,40$, respectively.
 The upper two curves (the dot for the jet with energy $E/\mu=40$,
 the solid for the jet with energy $E/\mu=10$)
 are for the contributions to the first order opacity (denoted as ``0+1").
 The long-dashed curve is for the self-quenching contribution (denoted as ``0")
 from the jet with $E/\mu=10$ while
 the short-dashed curve for the induced contribution (denoted as ``1")
 from the same jet at the first order opacity.
 The opacity is chosen as $L/\lambda=6$.}
 \end{figure}

 \begin{figure}
 \epsfxsize 80mm \epsfbox{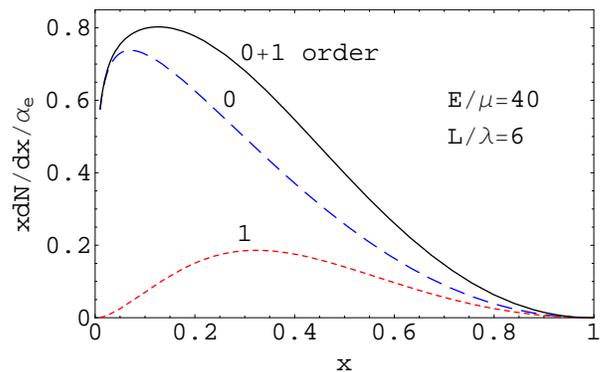}
 \vskip 0.2cm
 \caption{\label{fig3}
 (Color online) The radiation rate of the electromagnetic energy from a
 energy-given quark jet by radiating photons
 carrying energy fraction $x$ of the jet.}
 \end{figure}

To avoid the collinear divergences, in the following we take the
thermal mass of quark and gluon from ``hard thermal
loops"\cite{Braaten90} as
\begin{eqnarray}
    m^2&=&\frac{g_s^2 C_F T^2}{8}\, ,\qquad (C_F=\frac{4}{3})\, ,
  \label{m}\\
    \mu^2&=&(N_c+\frac{N_f}{2})\frac{g_s^2 T^2}{9}\, ,
  \label{mu}
\end{eqnarray}
where $N_c$ and $N_f$ are the number of the color and flavor of
quark, respectively. After taking into account the thermal mass of
quark, the collinear divergences in Eq.(\ref{f(xuy)}) disappear.
In following numerical calculation, we will choose $\mu=0.5$GeV
given by Eq.(\ref{mu}), the mean free path $\lambda=1fm$, and
consider the kinetic limits of the photon's transverse
momentum\cite{Wang01},
\begin{equation}
    \mu^2\leq{\bf k}_{\perp}^2\leq 4E^2x(1-x)\, .
\end{equation}

 \begin{figure}
 \epsfxsize 82mm \epsfbox{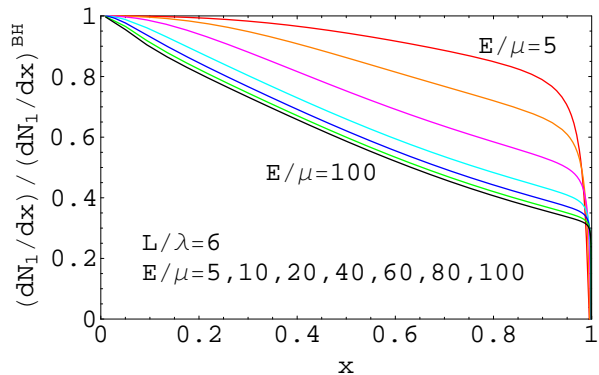}
 \vskip 0.2cm
 \caption{(Color online) The ratio between the rescattering contribution
 at the first order opacity and that of corresponding BH-limit for
 a quark jet with different energies.}
 \label{fig4}
 \end{figure}

Shown in Fig.\ref{fig2} is the production rates of the photon
emitted off two energy-given quark jets, respectively. The total
photon production rate (``0+1" order opacity) decrease with the
increasing photon energy ($xE$) and increases with the increasing
jet energy. The variation tendence of the total rate is the same as
AMY and TGJM's studies on bremsstrahlung contribution
\cite{AMY,TGJM}. Shown in Fig.\ref{fig3} is the radiation rate of
the electromagnetic energy from a energy-given quark jet by
radiating photons carrying energy fraction $x$ of the jet. The
variation tendence of the radiation rate is also similar to that of
AMY's studies on bremsstrahlung contribution. Of interest is that
our $``0"$ and $``1"$ order results can be directly compared with
the $``vaccum"$ and the $``induced"$ contributions in Zakharov's
studies, respectively. As $\mu=0.5$GeV is chosen in Eq.(\ref{mu}),
the jet energy shown in Fig. \ref{fig3} has a value $E=20$GeV same
as an example in Zakharov's studies \cite{Zakharov04}. The variation
tendence of our $``0"$ order contribution is similar to that of
$``vaccum"$ contribution in Zakharov's studies. However, the peak of
$``1"$ order correction contribution is at $x=0.3$ in our study
while the peak of $``induced"$ correction contribution is at $x=0.9$
for finite kinematic boundaries in Zakharov's studies. For a jet
created in $A+A$ collisions, multiple scattering induces photon
radiation as well as gluon radiation (jet quenching). Compared with
the case of $p+p$ collisions, induced photon radiation enhances the
photon yield while jet quenching decreases the photon yield from jet
fragmentation in $A+A$ collisions. If the former effect is estimated
smaller than the latter effect, the nuclear modification factor for
medium $p_T$ photon production $R_{AA}<1$ could be expected in $A+A$
collisions, close to experiment findings\cite{Isobe:2007ku}. Our
result for induced photon contribution smaller than that in
Zakharov's study may explain why a too larger enhancement factor of
photon production at high $p_T$ in $A+A$ is found in Zakharov's
study.

 \begin{figure}
 \epsfxsize 81.5mm \epsfbox{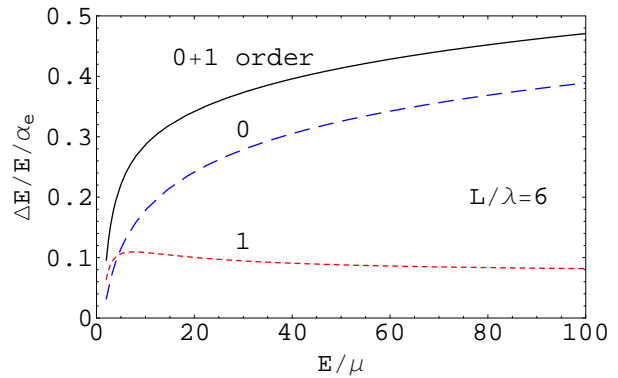}
 \vskip 0.2cm
 \caption{(Color online) The total electromagnetic energy loss
 as a function of the energy of a quark jet.}
 \label{fig5}
 \end{figure}

 \begin{figure}
 \epsfxsize 80.5mm \epsfbox{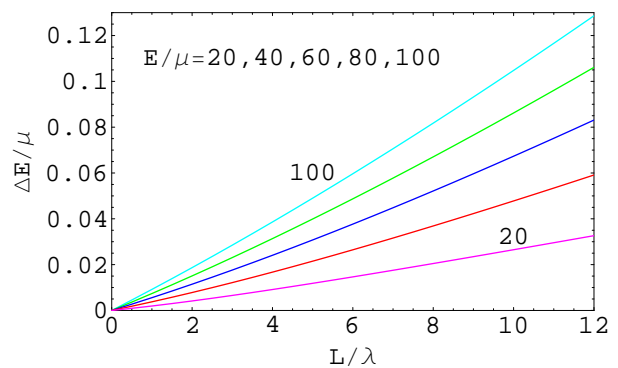}
 \vskip 0.2cm
 \caption{(Color online)
 The total electromagnetic energy loss of a quark jet ($E/\mu=5-100$)
 as a function of the thickness of the strong interaction medium.}
 \label{fig6}
 \end{figure}

It is the same as pointed out in Ref.\cite{Wang94}, from the
relative phase factor $\omega_0(z-z_0)$ in Eq.(\ref{R}) we define
the formation time of the photon radiation as
\begin{equation}
   \tau\equiv\frac{1}{\omega_0}=
   \frac{xp^+}{{\bf k}_{\perp}^2+x^2m^2}\, .
\label{tau}
\end{equation}
When $z-z_0\gg\tau$, the photon radiation reach the Bethe-Heitler
(BH) limit in which the intensity of the induced radiation is
additive in the number of rescattering. The BH-limit corresponds to
$I(\omega_0, L)=0$ in Eq.(\ref{lpm}). Shown in Fig.\ref{fig4} are
the ratio between the rescattering contribution at the first order
opacity and that of corresponding BH-limit for a quark jet with
different energies, $E/\mu=5-100$. From Fig.\ref{fig4} we see that,
at fixed jet energy, as $x$ approaches small value, $\tau$ becomes
small, the BH-limit is approached; at fixed $x$, as $E/\mu$ becomes
small, $\tau$ becomes also small (in Eq.(\ref{tau}), $p^+\sim 2E$),
the BH-limit is achieved again.

Integrating over $x$ in Fig.\ref{fig3} with different value of the
quark jet energy,
\begin{eqnarray}
    \Delta E/E&=&\int xdx\int d{\bf k}_\perp^2
    \frac{dN}{d{\bf k}_\perp^2dx}\, .
\label{eloss}
\end{eqnarray}
we obtain the total electromagnetic energy loss as a function of the
energy of a quark jet, shown in Fig.\ref{fig5}. For a fixed target
thickness $L/\lambda=6$, the self-quenching contribution increases
with the increasing jet energy while the rescattering contribution
($``1"$ order opacity correction) is almost a constant $\Delta
E/E/\alpha_e=0.1$ when $E/\mu>5GeV$. It show that for a high $p_T$
quark jet the leading contribution of the total induced
electromagnetic energy loss is proportional to the jet energy. Shown
in Fig.\ref{fig6} is the total electromagnetic energy loss of a
quark jet ($E/\mu=5-100$) as a function of the thickness of the
strong interaction medium. It is clear that the leading contribution
of energy loss by photon emission has a linear dependence on the
thickness of the nuclear target, which is different from the energy
loss by gluon radiation where $\Delta E$ quadratically depend on the
thickness $L$\cite{Gyulassy00,Wang01}. Similar results are also
observed in Ref.\cite{bwz-w}.

\section{The induced dilepton production}
What we study in above section is the real photon radiation when an
energetic parton jet produced in heavy ion collisions propagates
inside the strong interacting medium. If the radiative photon
induced by multiple rescattering in above process is virtual, the
virtual photon can decay into a lepton pair ($l\bar l$). In this
section we investigate the induced dilepton production.

The Feynman diagram for dilepton production induced by
self-quenching, single and double Born rescattering in the medium
is shown in Fig.\ref{fig6}. The amplitude for the dilepton
radiation reads
\begin{eqnarray}
    {\cal M}&=&C_\mu\frac{-ig^{\mu\nu}}{k^2}\bar u(l^+)(-ie\gamma_\nu)u(l^-)\, ,
  \\
    |{\cal M}|^2&=&\frac{e^2}{k^4}W^{\mu\nu}L_{\mu\nu}\, ,
\label{Md}
\end{eqnarray}
where
\begin{eqnarray}
    W^{\mu\nu}&=&C^\mu C^{\nu\ast},
  \\
    L_{\mu\nu}&=&4(l_\mu^+l_\nu^-+l_\mu^-l_\nu^+-l^+\cdot l^- g_{\mu\nu}),
\label{WL}
\end{eqnarray}
and $C^\mu$ is relate to the radiation amplitude for the virtual
photon.

 \begin{figure}
 \vskip -.5cm
 \epsfxsize 115mm \epsfbox{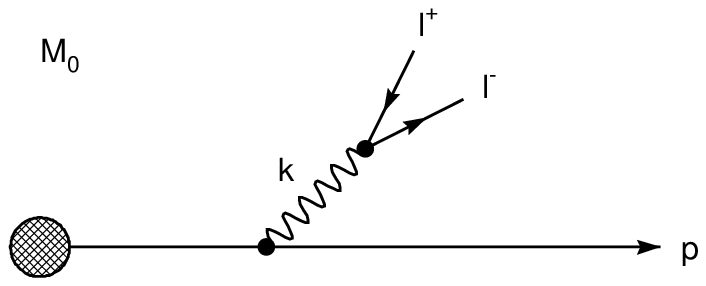}
 \vskip -1.5cm
 \epsfxsize 90mm \epsfbox{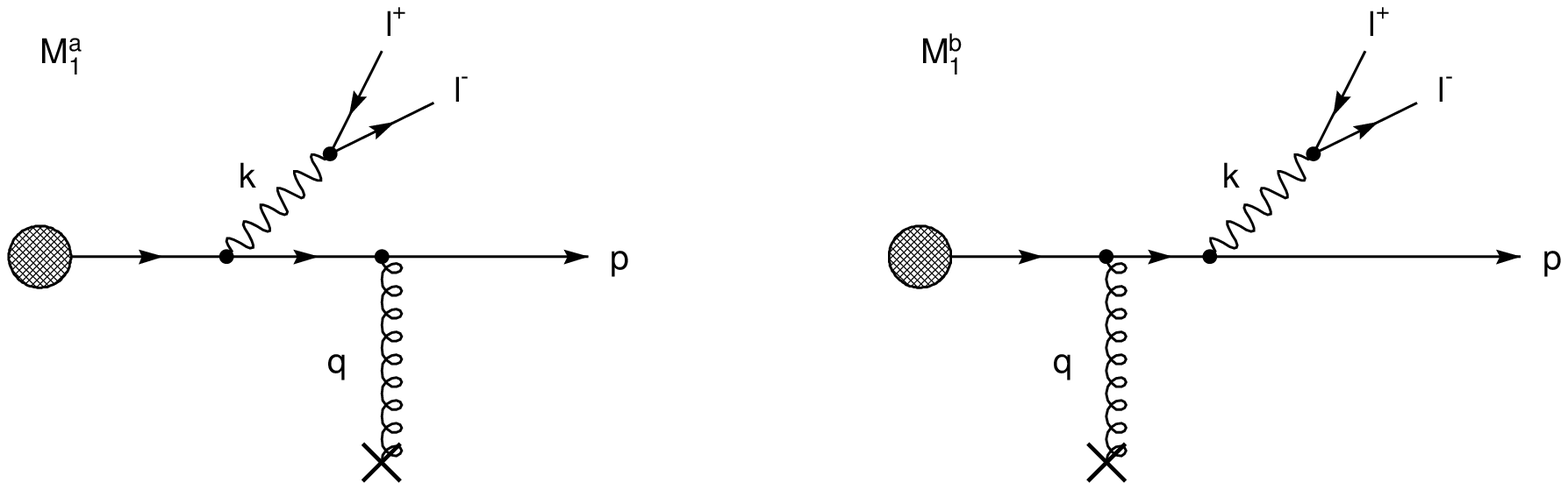}
 \vskip 0.1cm
 \epsfxsize 90mm \epsfbox{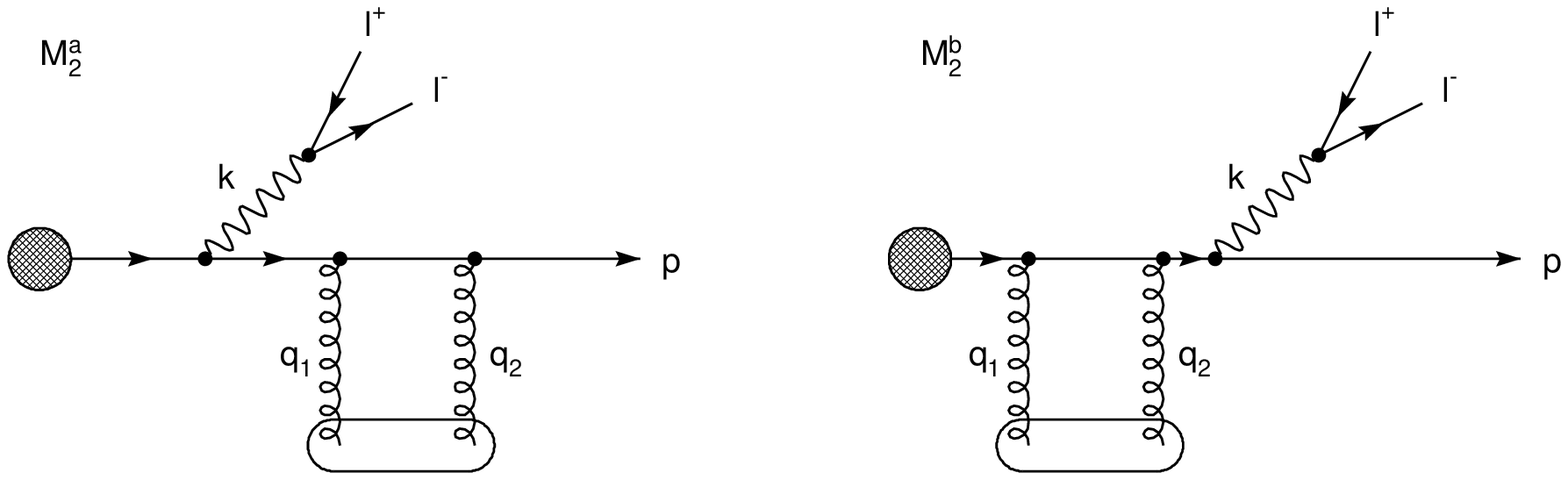}
 \vskip 0.2cm
 \caption{\label{fig7}
 \small Feynman diagram for dilepton production induced by
   self-quenching, single and double
   Born rescattering in the medium.}
 \end{figure}

The invariant-mass spectrum for dilepton production can be
expressed as
\begin{equation}
   \frac{dN}{dM^2}=\frac{\alpha_e}{3\pi M^2}(-g_{\mu\nu}W^{\mu\nu})
   \frac{d^3{\bf k}}{(2\pi)^32k^{0}}\, ,
\label{dNdM}
\end{equation}
where $k^2=M^2$, $M$ is the invariant mass of the dilepton.

For virtual massive photon field, we can choose its polarization
vector $\epsilon$ to satisfy $\epsilon\cdot k=0$. Correspondingly we
have\cite{Steman}
\begin{equation}
   \sum_{\lambda}\epsilon_\mu(\lambda)\epsilon_\nu^\ast(\lambda)
   =-g_{\mu\nu} +\frac{k_\mu k_\nu}{M^2}\, .
\label{pol}
\end{equation}
Then we obtain
\begin{eqnarray}
    -g_{\mu\nu}W^{\mu\nu}&=&\sum\left(\epsilon_\mu\epsilon_\nu^\ast
    C^\mu C^{\nu\ast}-\frac{k_\mu k_\nu^\ast C^\mu
    C^{\nu\ast}}{M^2}\right)
  \nonumber\\
    &=&\sum\left(|\epsilon_\mu C^\mu|^2-\frac{|k_\mu C^\mu|^2}{M^2}\right)
  \nonumber\\
    &=&\sum\left|\epsilon_\mu C^\mu\right|^2\, .
\label{gW}
\end{eqnarray}
In getting last equality the Ward Identity has been used.

As an approximation, we consider a massless quark jet and assume
$p^+\gg k^+\gg M$. From Eqs.(\ref{dNdM}) and (\ref{gW}) we see that
$\epsilon_\mu C^\mu$ is the radiation amplitude for a on-shell
virtual photon with mass $M$ reduced by rescattering. Its
calculation is very similar to the calculation of the radiation
amplitude for a massless real photon in Sec.II, the only difference
is that the light-cone 4-momentum in Eq.(\ref{k}) should be replaced
by
\begin{equation}
    k=\left[xp^+, \frac{{\bf k}^2_{\perp}+M^2}{xp^+},
    {\bf k}_{\perp}\right]\, .
\label{kM}
\end{equation}
For self-quenching, single and double Born rescattering, the
radiation amplitude of the virtual photon with mass $M$ can be
deduced as
\begin{equation}
    {\cal R'}_0=2(1-x)g {\bf\epsilon}_{\perp}\cdot{\bf B'}\, ,
\label{R'0}
\end{equation}
\begin{eqnarray}
    {\cal R'}_1 &=& -2ig(1-x)
  \nonumber\\
    &\times&{\bf \epsilon}_{\perp}\cdot\sum_{j=1}^N
    [{\bf B'}-({\bf B'}-{\bf C'}) e^{i\omega'_0(z_j-z_0)}]\, ,
\label{R'1}
\end{eqnarray}
\begin{equation}
    {\cal R'}_2^B=-(1-x)gN {\bf\epsilon}_{\perp}\cdot{\bf B'}\, ,
\label{R'2}
\end{equation}
where
\begin{equation}
   {\bf B'}=\frac{{\bf k}_{\perp}}{{\bf k}_{\perp}^2+M^2}\, ,
   \quad{\bf C'}=\frac{{\bf k}_{\perp}-x{\bf q}_{\perp}}
   {({\bf k}_{\perp}-x{\bf q}_{\perp})^2+M^2}\, ,
\label{B'C'}
\end{equation}
and
\begin{equation}
  \omega'_0=\frac{{\bf k}_{\perp}^2+M^2}{xp^+}\, .
\label{om'}
\end{equation}
Correspondingly, the square of the radiation amplitude for the
virtual photon to the first order in opacity expansion can be
expressed as
\begin{eqnarray}
    &&-g_{\mu\nu}W^{\mu\nu}=|{\cal R'}_0|^2+\frac{L}{\lambda}
    \int\frac{\mu^2d^2{\bf q}_{\perp}} {\pi({\bf q}^2_{\perp}+\mu^2)^2}
  \nonumber\\
    &&\qquad\times\int dz_0 dz
     \rho(z_0,z)[|{\cal R'}_1|^2+ 2Re({\cal R'}_2^B{\cal R'}_0^{*})]\, ,
\label{amplitude}
\end{eqnarray}
where
\begin{eqnarray}
    |{\cal R'}_1|^2&+&2Re({\cal R'}_2^B{\cal R'}_0^{*})
    =4g^2(1-x)^2[({\bf B'}-{\bf C'})^2
  \nonumber\\
     &&\qquad-2\cos(\omega'_0(z-z_0))
     ({\bf B'}^2-{\bf B'}\cdot{\bf C'})]\, .
\label{R'}
\end{eqnarray}
It is the same as in Eq.(\ref{R}), the term with
$\cos(\omega'_0(z-z_0))$ represents the destructive interference
resulting from the Abelian LPM effect\cite{LPM}.

Substituting Eq.(\ref{amplitude}) into Eq.(\ref{dNdM}), we can
obtain the invariant-mass spectrum arising from self-quenching at
zero order in opacity,
\begin{equation}
   \frac{dN_{0}}{dxdydw}=
   \frac{4\alpha_e^2}{27\pi^2}\frac{(1-x)^2}{xw}
   \frac{y}{(y+w)^2}\, ,
\label{dN0dxdydw}
\end{equation}
and the invariant-mass spectrum arising from single and double
Born rescattering at first order in opacity,
\begin{eqnarray}
    \frac{dN_{1}}{dxdydw}&=&
    \frac{4\alpha_e^2}{27\pi^2}\frac{(1-x)^2}{xw}\frac{L}{\lambda}
  \nonumber\\
    &\times &\int du\frac1{(1+u)^2}f(x,u,y,w)\, ,
\label{dN1dxdydw}
\end{eqnarray}
where function
\begin{eqnarray}
    &&f(x,u,y,w)=\frac{(y+x^2u)(y+w+x^2u)-4x^2uy}
    {\sqrt{((y+x^2u+w)^2-4x^2uy)^3}}
  \nonumber\\
    &&\qquad-\frac{y}{(y+w)^2}+(1-I(\omega'_0, L))
    \frac{1}{y+w}
  \nonumber\\
    &&\qquad\times\left(\frac{y-w}{y+w}
    -\frac{y-w-x^2u}{\sqrt{(y+x^2u+w)^2-4x^2uy}} \right)\, ,
\label{f(xuyw)}
\end{eqnarray}
and
\begin{equation}
   u=|{\bf q}_{\perp}|^2/\mu^2\, ,\quad y=|{\bf k}_{\perp}|^2/\mu^2\, ,
   \quad w=M^2/\mu^2\, .
\end{equation}

 \begin{figure}
 \epsfxsize 80mm \epsfbox{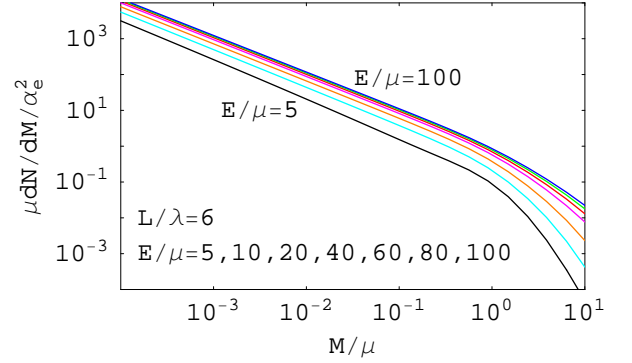}
 \vskip 0.2cm
 \caption{\label{fig8}
 \small (Color online) The invariant-mass spectra (``0+1" order) of the dilepton emitted off
 quark jets with different energies, $E/\mu=5,10,20,40,60,80,100$,
 passing through the medium with the thickness, $L/\lambda=6$. }
 \end{figure}

\begin{figure}
 \epsfxsize 80mm \epsfbox{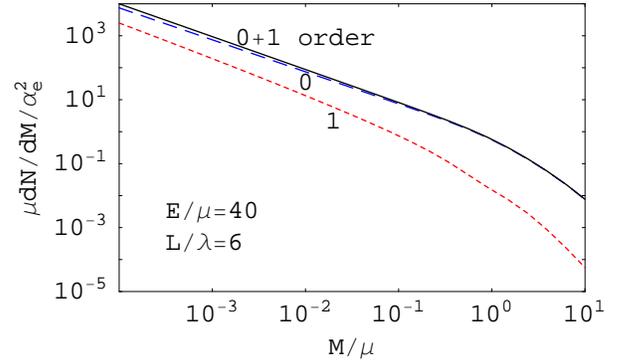}
 \vskip 0.2cm
 \caption{\label{fig9}
 \small (Color online) The comparison of the induced contribution (denoted as ``1 order")
 and the self-quenching contribution (denoted as ``0") of the invariant-mass spectra
 of the dilepton emitted off an energy-given
 quark jet, $E/\mu=40$, passing through the medium with the thickness, $L/\lambda=6$.
}
 \end{figure}

 \begin{figure}
 \epsfxsize 80mm \epsfbox{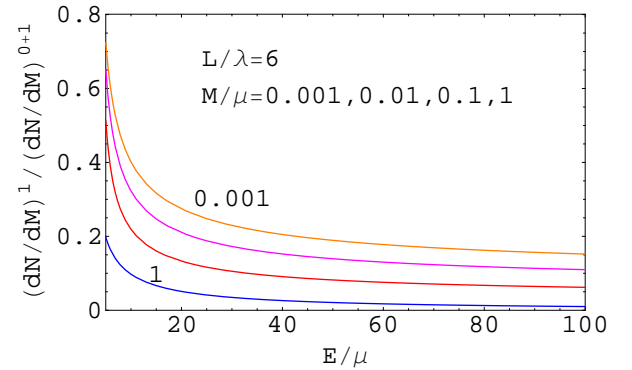}
 \vskip 0.2cm
 \caption{\label{fig10}
 \small (Color online) The ratio of the invariant-mass spectra between 1-order and (0+1)-order in opacity
         expansion. The invariant-mass is
         chosen as
         $M/\mu=0.001,0.01,0.1,1$, respectively. }
 \end{figure}

Shown in Fig.\ref{fig8} is the total invariant-mass spectra (``0+1"
order opacity) of the dilepton emitted off quark jets with different
energies, $E/\mu=5,10,20,40,60,80,100$, passing through a target
with the thickness, $L/\lambda=6$. The invariant-mass spectrum
decreases with the increasing invariant-mass while increases with
the increasing jet energy. Shown in Fig.\ref{fig9} is the comparison
of the induced contribution (``1 order" opacity) and the
self-quenching contribution (``0" opacity) of the invariant-mass
spectra of the dilepton emitted off an energy-given quark jet. For a
jet with energy $E/\mu=40$, the rescattering contribution is very
small and can be neglected in the large invariant-mass region
(especially when $M/\mu>1$). In order to know clearly how much the
rescattering contribution is in the total contribution, we plot
Fig.\ref{fig10} to check the ratio of the invariant-mass spectra as
a function of the jet energy between 1-order and (0+1)-order in
opacity expansion. The invariant-mass is chosen as
$M/\mu=0.001,0.01,0.1,1$, respectively. The plot demonstrates that
the dilepton production induced by rescattering is important for
small value of the invariant-mass and also the jet energy. In
addition, the contribution fraction by rescattering is found to be
nearly a constant when $E/\mu>30$.

\section{Conclusion}
In this paper we apply GLV's opacity expansion technique to
investigate the induced photon radiation and dilepton production
in strong interacting medium. The real photon radiation and the
dilepton invariant-mass spectrums are presented to the first order
in opacity expansion. Our study focuses on the electromagnetic
bremsstrahlung of an energy-given quark jet passing a
thickness-given QGP system and the analytical derivation is combined
with numerical calculation.

We show that for real
photon radiation due to Abelian LPM effect the parton
jet's energy loss has a linear dependence on the thickness of the
targets instead of quadratic dependence arising from non-Abelian
LPM effect for gluon radiation. And the leading contribution of
the total induced electromagnetic energy loss of a fast quark
is proportional to the jet energy.

The dilepton production by the decay from the virtual photon off a quark jet
induced by multiple rescattering in medium has been investigated.
It is shown that the
invariant-mass spectrum decreases with the increasing
invariant-mass while increases with the increasing jet energy. The
rescattering contribution is important for small value of the
invariant-mass and also for the not so fast jet. In addition, the
contribution fraction by rescattering is found to be nearly a
constant when the quark jet energy $E/\mu>30$.

Of course, the study presented here is just the first step for
photon and dilepton production in medium. To make a complete study
of photon and dilepton production in $A+A$ collisions many other
contributions should be taken into account, such as the prompt
photon produced by the initial hard scattering, fragmentated photons
by jet fragmentation, jet-photon conversion\cite{Fries:2002kt}, et
al. A systematical study will also include other nuclear effects
such as Cronin effect, (anti-)shadowing effect and final jet
quenching effect for a jet fragmentated into a photon, just like our
current studies\cite{zoww07,zhz-qm08} for high $p_T$ hadrons in
$A+A$ collisions. This kind of detailed calculating and analyzing
for large transverse momentum photon production will be left for a
future's study.

\acknowledgments

The authors are very grateful to Xin-Nian Wang for helpful
discussions. This work was supported by MOE of China under Projects
No. IRT0624, No. NCET-04-0744 and No. SRFDP-20040511005, and by NSFC
of China under Projects No. 10440420018, No. 10475031 and No.
10635020.

\end{document}